\documentstyle[preprint,aps]{revtex}
\begin{document}

\newcommand\beq{\begin{equation}}
\newcommand\eeq{\end{equation}}
\newcommand\bea{\begin{eqnarray}}
\newcommand\eea{\end{eqnarray}}
\newcommand\bc{\begin{center}}
\newcommand\ec{\end{center}}
\newcommand\ti{\tilde}

\draft
\title{Twocomponent Fermi vapour in a 2D rotating trap }

\author{Sankalpa Ghosh $^1$, M.V.N.Murthy$^1$ and Subhasis Sinha$^2$} 
\address{1. The
Institute of Mathematical Sciences, Chennai 600113, India} 
\address{2. Laboratoire Kastler-Brossel, Ecole Normale Sup\'erieure,
24 rue Lhomond 75 231 Paris Cedex 05. France}
\maketitle

\begin{abstract}

An exactly solvable model of two-component interacting Fermi vapour in two
dimension within Thomas Fermi approach has been proposed. We assume a
realistic off-diagonal s-wave interaction between fermions in the two
hyperfine states. The interaction is taken to be given by a screened
Coulomb interaction which is both relevant and is analytically tractable. 
There are two distinct limits in the case of trapped fermionic systems: 
When the rotating frequency is less than the trap frequency a Thomas-Fermi
approach may be used for a reasonably large number of particles. In the
case of rapidly rotating fermions, an alternative variational approach is
used, where the local density approximation may be applied after
projecting the
system on to the lowest degenerate level. Analytic expressions for the
relevant physical quantities are obtained and their significance is
discussed in both these cases. 

\end{abstract}

%\tightenlines
\newpage
\section{Introduction}

The experimental observation of Bose-Einstein Condensation(BEC) in
extremely cold dilute gas of atoms has generated a lot of activity both on
experimental as well as theoretical fronts on the properties of trapped
dilute gases near the quantum degeneracy limit. There is now a renewed
focus on the properties of trapped dilute gas of fermionic atoms at low
temperatures. Magneto-optical confinement of fermionic gases has been
reported for $^6 Li$ \cite{WIM} and $^{40}K$ \cite{FSC}. A recent
significant success in this direction is the experimental observation of
quantum degeneracy in a dilute gas of trapped fermionic
atoms\cite{Demarco}. They have realised the magneto-optical trapping of
$^{40}K$ atoms in two different hyperfine states corresponding to
$|F=9/2,m_F=9/2>$ and $|F=9/2,m_F=7/2>$ and finally reaching a single
component Fermi vapour by selective removal of atoms in one of the
hyperfine states. Amoruso ${\it et. al.}$ \cite{Tosi} have studied the
ground state and small amplitude excitations of such two component
degenerate Fermi vapours placing them in spherically symmetric trap and
discuss the effect of interaction on dissipative hydrodynamic processes in
them. Salassnich ${\it et. al.}$\cite{salasnich} have studied the
thermodynamic properties of such multicomponent fermi vapours
theoretically and have pointed out that such systems can show phase
separation and as well as shell effects.

While the effective interaction between $^{40}K$ atoms in two different
hyperfine states is repulsive a negative scattering length for $^6 Li$
atoms holds promise of achieving a superfluid state in a mixture of $^{6}
Li$ atoms prepared in two hyperfine states \cite{Stoof}. This has
motivated a number of theoretical papers. Recently collective excitations
of the system in the normal phase\cite{brunn} and in the superfluid
phase\cite{baranov} have been investigated. Ghosh and Sinha\cite{Tarun}
have studied \cite{Tarun} the low lying collective excitations of Fermi
and Bose systems in an anisotropic trap. 

Especially relevant to this paper is the recent interest in the properties
of rotating bose condensates\cite{Butts} and the experimental observation
of vortices in stirred BEC of $^{87} Rb$\cite{expt}. The properties of
ground and low excited states of a rotating and weakly interacting
Bose-Einstein condensate in a harmonic trap has been investigated recently
in various limits\cite{Mot1,kavoulakis,Castin}.  Benakli ${\it et. al.}$
\cite{shenoy} have studied the properties of macroscopic angular momentum
state of a rotating BEC in a toroidal trap and identified $l$-dependent
density profile as a signature of condensate rotation and superfluidity. 
Another very interesting theoretical issue related to the rotating BEC is
the exact mapping of the problem in the limit of weak interaction to that
of Lowest Landau Level problem in Quantum Hall effect \cite{LLL}.
Recently, experiments\cite{MIT} has been able to trace a collective state
of 160 vortices in a BEC consisting  of $Na$ atoms by imparting a large 
angular momentum to the system by rotating it.
. This experiment is a
major step towards realising this weakly interacting regime where the LLL
mapping is useful.  Theoretical model of such vortex lattice state has
been discussed  by Ho by invoking the fact that this system has a
strong resemblance with the Quantum Hall systems.(\cite{Ho1}). 

It is therefore quite natural to extend the analysis to Fermi vapours
also. Ho and Ciobanu\cite{Ho} have recently examined theoretically the
nature the ground state of a rapidly rotating trapped Fermi gas in two and
three dimensions in the noninteracting limit. They show that the density
profile acquires features reflecting the underlying Landau level like
energy spectrum when the frequency of rotation approaches the bare
confinement frequency. Properties of the spectrum of such a system had
been earlier investigated in detail by Bhaduri et al\cite{bhaduri}. 

A simple and straight forward way of studying the ground state properties
of such interacting many-body system is through the Thomas-Fermi(TF)
approximation. Butts and Rokshar\cite{rokshar} studied the momentum and
spatial distribution of the noninteracting system and identify the region
in which TF approximation is valid. 
The ground state properties and the addition
energy spectra of a two-dimensional interacting Fermi system has been
studied by Sinha et al\cite{subhasis} within Thomas Fermi approximation
using an exactly solvable model interaction. A similar exactly solvable
TF
model has been discussed for a two dimensional parabolic quantum dot in a
weak magnetic field by R. Pino (\cite{Pino}). 

In this paper we analyse a multicomponent Fermi vapour confined in a
rotating trap in two dimension within the Thomas Fermi approximation and
with an effective interaction between the fermions. If the confinement
along the axis of rotation ($z$- axis) is relatively much stronger, then
only a few lowest energy subbands along that direction get populated. In
such a situation the system becomes effectively a multicomponent two
dimensional Fermi gas, the number of components being the number of
subbands being occupied. In particular when one uses optical traps such a
system is experimentally realisable \cite{DMS}.  In the case of rotating
Bose gas such two-dimensional systems have been extensively studied
\cite{Castin}, but similar studies for their fermionic counterparts is
still
far from complete.  The effective interaction is given by a screened
Coulomb potential with a finite range. We show that within the TF
approximation the system may be solved exactly and discuss the consequent
results. We follow closely the formalism developed by Gallego and Das
Gupta\cite{gdg} who developed the TF method for the rotating nuclei. An
attractive feature of the TF method is the ease with which non-trivial
many-body solutions can be obtained even when they are interacting.  As we
have shown in a preceeding work \cite{ghosh}, in some cases the results
may be obtained purely by analytical methods. In this paper we discuss a
model interaction which is relevant to physical systems. Appropriate
limits can reproduce the results already obtained in ref. \cite{ghosh}.
Though our discussion is centered around rotating fermions in two
dimension it is equally applicable to systems like quantum dot in a
parabolic confinement with\cite{Pino} or without a magnetic
field\cite{subhasis}. 

The paper is organised as follows. In section II we outline the derivation
of the Thomas Fermi energy functional for a system of interacting rotating
fermions in parabolic confinement and discuss our choice of effective
interaction. In section III we apply the formalism to a two-component
Fermi gas when the rotational frequency is less than the confinement
frequency and obtain analytic expressions for various quantities of
physical interest.  We also present numerical results for the interaction
energy, the rotational energy and the spatial density in a given angular
momentum sector for a given number of particles.  Furthermore we discuss
the extreme quantum limit where a rapidly rotating Fermi gas can be mapped
exactly to the problem of a charged particle in uniform transverse
magnetic field in 2 -dimension.  Exploiting this similarity we have
analysed the many particle spectrum through a variational technique which
is an improvement over the usual semiclassical method.  The last section
contains a discussion and summary of the results.

\section{Thomas Fermi Energy functional}

We derive the Thomas-Fermi(TF) energy functional for a confined two
dimensional rotating fermi gas starting from the microscopic Hamiltonian.
We shall follow the approach outlined by Gallego and Das gupta \cite{gdg}
who derived the TF density functional for rotating nuclei, but applied to
a two dimensional system. 

The microscopic Hamiltonian of a two-dimensional rotating Fermi gas 
of $N$ particles confined in a harmonic potential is given by,
\bea
H &=& \sum_{i=1}^N \frac{{\bf p}_i^2}{2m} + \sum_{i<j}V(\vec 
r_i,\vec r_j) +\sum \frac{1}{2}m \omega_0^2 r_i^2 \nonumber \\
& &\mbox{}-\sum m\omega[{\bf r_i} \times {\bf p_i}], 
\label{MBH} 
\eea
where $V(\vec r_i, \vec r_j)$ denotes the two body interaction between 
fermions, $\omega_0$ and $\omega$ denote the confinement and rotational 
frequencies respectively.

We want to minimise the energy of this system subject to the
constraints
\bea 
\int d \vec r \rho(\vec r) &=& N,  \label{cons1} \\
\int d \vec r d\vec p f(\vec r,\vec p)(\vec r \times \vec p) &=& L, 
\label{cons2} 
\eea  
where $N$ and $L$ are total particle number and the total Angular
momentum, $f(\vec r,\vec p)$ is the semiclassical phase space density
which we shall define later.  The configuration space density $\rho$ is
obtained by integrating the phase space density over the momentum space. 
 
Using Lagrange multiplier $\mu$ and $\omega$ for each constraint
respectively and regrouping the momentum dependent term we obtain the
following expression for the energy functional 
\bea 
E[\rho] &=& \int d\vec r d\vec pf (\vec r, \vec p)[\frac{p^2}{2m}
-\omega(xp_y - yp_x)] \nonumber \\ 
& & \mbox{} +  \int d{\vec r}[V(\vec r) - \mu] \rho(\vec r)+ \mu N
+ \omega L \label{enfun} 
\eea
where the mean-field one body interaction $V(r)$ is 
\beq 
V(r) = \frac{1}{2}m \omega_0^2 r^2 + V_{H}(\vec r).\eeq
Note that the formalism itself is valid for arbitrary confinement
potentials  as long as
the meanfield has a smooth behaviour.  However in the later sections we
restrict our analysis to parabolic confinement.

The Hartree term is given by,
\beq 
V_{H}(\vec r)=  \int d^2r V(\vec r, \vec r~') \rho(\vec r~'). 
\eeq
We ignore the exchange correction as it is in general subdominant 
compared to the Hartree term.

The semiclassical phase space density is given by 
\beq 
f(\vec r, \vec p) = \frac{2}{ (2\pi \hbar)^2}\Theta(\mu - \epsilon(\vec p,
\vec r)), 
\eeq 
where we have accounted for a factor of 2 due to spin degeneracy and 
\beq 
\epsilon(\vec p, \vec r) =\frac{p^2}{2m} +V(\vec r)
- \omega(xp_y - yp_x)
\eeq
denotes the energy density in phase space. We may rewrite this by
completing the square as,
\beq 
\epsilon(\vec p, \vec r) =\frac{1}{2m}[(p_x+m\omega y)^2 + (p_y -m\omega
x)^2] +V(\vec r)- \frac{1}{2}m\omega^2 r^2.
\eeq
The shifted momentum simply indicates that the center of the Fermi sphere
at any given point $\vec r$ in the rotating frame is displaced from the
usual $\vec p =0$. This shift is of no relevance within the classical TF
approximation except for the appearance of the extra centrifugal term in
the energy functional.

The TF energy functional is therefore given by,
\beq 
E[\rho] = \int d^2 r~[\frac{\pi \hbar ^2 \rho^2}{2m} + \frac{1}{2}m
\Omega r^2 + V_{H} - \mu] \rho(\vec r) + \mu N + \omega L 
\label{E}
\eeq
where
\beq 
\Omega^2~=~\omega_0^2 - \omega^2
\eeq
denotes the effective frequency in the presence of rotations. 
The TF equation for the spatial density is obtained by a variation of the
energy functional, namely,
\beq 
\frac{\pi \hbar^2 \rho}{m} + \frac{1}{2}m\Omega^2 r^2 
+ V_{H}  =\mu 
\label{TFE} 
\eeq
To obtain the ground state density in TF approximation one has to solve
the above equation self-consistently with the boundary condition 
that the density vanishes beyound the classical turning point $r_0$, that 
is 
\beq \rho(r) = 0 , ~~~r \ge r_0. \eeq
Note that in the absence of rotations $\Omega=\omega_0$ and the solution
of the above equation describes the ground state in the absence of
rotations which has been analysed in detail before\cite{subhasis}.
We may remark that in general the above equation is difficult to solve
analytically except in specific cases which we shall discuss later.

The angular momentum carried by the system is given by,
\beq 
L = \int d^2 r d^2 p f (\vec r, \vec p)(xp_y - yp_x) 
=m\omega\int d^{2}r \rho(\vec r) r^2 
\eeq 
which is simply the classical expression for the angular momentum.
We may therefore identify the Lagranges multiplier $\omega$ with the
angular frequency.

We may also define the related energy functionals. The free energy of the
system is given by,
\beq 
F = E - \mu N - \omega L 
\eeq
The TF equation (\ref{TFE}) may also be obtained just by varying the free 
energy. We also define the energy in the rotating frame through 
\beq 
E = E' + \omega L, 
\eeq
where E is the total energy in the static frame.  With the help of eq.
(\ref{TFE}) it can be shown that irrespective of the nature of the
interaction $E'$ is given by,
\beq 
E' = \frac{1}{2}\mu N + \frac{1}{4} m \Omega^2 \int d^2r~ r^2 \rho.
\eeq
Note that it is necessary that we choose $\omega \le \omega_0$ as
otherwise the system becomes unstable. 

This concludes the basic TF formalism applied to rotating fermionic
systems. In the following section we solve the TF equation for a specific
case. In an earlier brief note (\cite{ghosh}) we considered some examples
of model interactions for which the TF equations are exactly solvable for
a circularly symmetric density profile. In here we extend our calculation
to the case of effective (screened) Coulomb interaction. Depending on the
range (which is a free parameter in our calculation) the interaction
yields both logarithmic as well as the contact interaction.  

The Screened Coulomb interaction for a two dimensional system is given by
\beq 
V(\vec r)=\frac{g}{2 \pi}\int \frac{e^{i \vec k . \vec r}
d \vec k}{k^2 +\delta^2}, 
\eeq
where $g$ is the coupling constant and the $\delta$ denotes the relevant 
mass scale which determines the range of interaction. The choice of the 
effective interaction is dictated not only by the fact that it is 
tractable, but also by the fact that it is a good approximation to the 
realistic interaction between charged fermions in two dimensions.
Given the very general form we have used in terms of the coupling 
constant $g$ and the range $b$  even for the neutral atomic (Fermi) vapours 
the consideration of  such finite range interaction is quite
important(\cite{maxim}). 
It can be checked that the above integral can easily written as 
\beq
V(r)=gK_0(\frac{r}{b}), 
\eeq
where $K_0$ is the modified Bessel function and $b =\frac{1}{\delta}$ 
denotes the range of the interaction. 

When the range is very small then the given form of $V(\vec r)$ behaves
almost like contact interaction and in the limit of $b\rightarrow\infty$
the interaction goes over to the logarithmic interaction which is Coulomb
interaction in an ideal two-dimensional system. These can be checked from
the following limiting behaviour of the modified bessel function $K_0(z)$
\cite{AS}:

$b \rightarrow \infty \Rightarrow \frac{r}{b} \rightarrow 0, $
\beq 
K_0(\frac{r}{b}) \sim  -\log (\frac{r}{b}) 
\eeq 
and

$b \rightarrow 0 \Rightarrow \frac{r}{b} \rightarrow \infty$
\beq 
K_{0}(\frac{r}{b}) \rightarrow 2 \pi b^2\delta^2(\vec r)
\eeq

Hence in the two extreme limits we obtain the logarithmic and the contact
like behaviour.  In  Fig. 1 we have plotted for comaprisons three
different potentials $V(x)$ as a function of$x$ - namely the Coulomb in
two dimension $1/x$, the negative of logarithmic interaction $log|x|$, and
the screened Coulomb $K_0(x)$. All distances are measured in units  of
length is defined by $l_0 = \sqrt{\frac{\hbar}{m \omega_0}}$
. For the
purpose of comparison we have set the range $\ti{b}=1$. It is easy to see
that at long distance screened Coulomb falls faster than the unscreened
Coulomb potential whereas at the sort distance it goes like -$log|x|$.
We also note that for comparison we should only look at the 
portion of the plot
of $-log|x|$  upto $x=1$ where the potential vanishes. 

The TF analysis in these two limits has already been done in the previous
paper\cite{ghosh}. In the following section we solve the TF equations for
a general screened Coulomb interaction and discuss the results. The only
approximation which has gone into obtaining the solutions is the
assumption that the density is circularly symmetric.

\section{Two-component Fermi gas with effective screened Coulomb
Interaction}

For a two component system the coupling constant can be written in a $ 2
\times 2$ matrix form to take care of inter and intra component
interaction. We assume that such interaction strength matrix is symmetric
in its off-diagonal components ${\it i.e.,}$ $g_{12}=g_{21}=g$. Due to the
anti-symmetry of the many body wave function the fermions occupying the
same hyperfine state do not have the s-wave interaction but interact
through the much weaker p-wave interaction. To obtain the leading order
results, we neglect this. Therefore we set $g_{11}=g_{22}=0$. 
 
The Thomas Fermi free energy functional of such  two component Fermi 
vapour is therefore given by
\bea 
F[ \rho_1 ,\rho_2] &=& \int d{\bf r} [\frac{\pi
\hbar^2}{m}(\rho_1^2 +\rho_2^2) + \frac{1}{2}m\Omega^2 r^2(\rho_1
+ \rho_2) \nonumber \\ & & \mbox{} + 
 \rho_{1}(\vec r)\int d{\bf r'} V_{1
2}(|\vec r - \vec r'|)\rho_{2}(\vec r')] -\mu_1 N_1 - \mu_2
N_2,
\label{FREEENG} 
\eea
where $\rho_1, \rho_2$ denotes the densities of the two components while 
$N_1$ and $N_2$ denote the populations of the two hyperfine states which 
is kept fixed during minimisation. Furthermore, we have
\beq 
V_{12}(|\vec r - \vec r'|)  =
\frac{g_{1
2}}{2 \pi}
\int \frac{e^{i \vec k . (\vec r - \vec r')}
d \vec k}{k^2 +\delta^2}
\eeq 
for the interaction matrix.

In what follows we use the dimensionless variables (denoted as
variables with $\tilde{}$ on top). The quantities with energy dimensions are
measured in
units of $\hbar \omega_0$ and the length is measured in units of $l_0 =
\sqrt{\frac{\hbar}{m \omega_0}}$. 
After this  rescaling the free energy functional may be written as,
\beq 
\ti{F}[\ti{\rho}_1,\ti{\rho}_2] = \int d{\vec x}[\pi(\ti{\rho}_1^2 +
\ti{\rho}_2^2) +
\frac{\ti{\Omega}^2 x^2 (\ti{\rho}_1+\ti{\rho}_2)}{2} + \ti{\rho}_{1} \int
d{\vec x}'V_{12}(x,x')\ti{\rho}_{2}(x')] - \ti{\mu}_1N_1 - \ti{\mu}_2 N_2 ,
\eeq
where scaled variables are given by
\bea
\ti{\mu}_{\alpha} &=& \frac{\mu_{\alpha}}{\hbar \omega_0}  \nonumber \\
\ti{\Omega}^2 &= & \frac{\Omega^2}{\omega_0^2} \nonumber \\
\ti{\rho}_{\alpha} &=& \rho_{\alpha}l_0^2 \nonumber \\
 x &=& \frac{r}{l_0} 
\eea
 
Minimising the free energy functional, we get the equations for the TF 
densities,
 
\beq 
2\pi\ti{\rho}_1 = \ti{\mu}_1 - \ti{V}_{H1} - \frac{1}{2}\ti{\Omega}^2 x^2,
\label{tf1}
\eeq 
\beq 
2\pi\ti{\rho}_2 = \ti{\mu}_2 - \ti{V}_{H2} - \frac{1}{2}\ti{\Omega}^2 x^2, 
\label{tf2} 
\eeq
where 
\beq 
\ti{V}_{H \alpha}(x) = \frac{1}{\hbar\omega_0}\int d \vec x' V_{\alpha 
\beta}(|\vec x - \vec
x'|)\ti{\rho}_{\beta}(\vec x')
\eeq

By definition the Hartree potential obeys the following set of
Poisson's equations,
\beq
\nabla^2 \ti{V}_{H1}=- 2 \pi \ti{g} \ti{\rho}_2 +
\frac{\ti{V}_{H1}(x)}{\ti{b}^2},
\label{pot1} 
\eeq
and
\beq 
\nabla^2 \ti{V}_{H2}=- 2\pi \ti{g} \ti{\rho}_1 +
\frac{\ti{V}_{H2}(x)}{\ti{b}^2},
\label{pot2} 
\eeq
where the coupling $\ti{g} = \frac{g}{\hbar \omega_0}$ and the range
parameter
is now given by $\ti{b}=\frac{b}{l_0}$ and, $\ti{\rho}_{1,2}$ is the
density in
dimensionless units ($\ti{\rho}=\rho(\vec r)l_0^2$)

Substituing  Eqs. (\ref{pot1},\ref{pot2}) in eqs. (\ref{tf1},\ref{tf2}) 
we get following  equations for the two densities,
\beq 
2 \pi \nabla^2 \ti{\rho}_1 + 2 \ti{\Omega}^2 -2 \pi \ti{g}\ti{\rho}_2 
+\frac{1}{\ti{b}^2}(\ti{\mu}_1 - 2\pi\ti{\rho}_1-\frac{1}{2}x^2
\ti{\Omega}^2) = 0 
\label{rho1} 
\eeq
and
\beq 
2 \pi \nabla^2 \ti{\rho}_2 + 2 \ti{\Omega}^2 -2 \pi \ti{g}\ti{\rho}_1 
+\frac{1}{\ti{b}^2}(\ti{\mu}_2 - 2\pi\ti{\rho}_2-\frac{1}{2}x^2
\ti{\Omega}^2) = 0 .
\label{rho2} 
\eeq
The eqs. (\ref{rho1}) and (\ref{rho2}) may be solved subject to the
usual boundary conditions 
\bea
\ti{\rho}_1 &=& 0 ~at~ x \ge x_1 \\
\ti{\rho}_2 &=& 0 ~at~  x \ge x_2 
\eea
At this stage it is useful to define the following quantities 
\bea
\ti{\rho}&=& \ti{\rho}_1 + \ti{\rho}_2,\\
\ti{s} &=& \ti{\rho}_1-\ti{\rho}_2 ,\\
\ti{\mu}^{\pm}&=&(\ti{\mu}_1 \pm \ti{\mu}_2), 
\eea
where $\ti{\rho}$ denotes the total density, and $\ti{s}$ denotes the spin
density. It is easy to see that in terms of the density and the spin
density the coupled equations for the two components are diagonal.  In
this paper we consider the case when the system is unpolarised, that is
when the spin density is zero. The case of partially polarised system will
be discussed in detail in a forthcoming publication\cite{future}. 

Before proceeding further we note that, since $\Omega^2 =\omega_0^2 -
\omega^2$, the system becomes unstable when the rotational frequency
approaches the confinement frequency. TF method can be applied abinitio
only to the case when the rotational frequency is less than the
confinement frequency. In the rapidly rotating case many levels collapse
into the lowest one reminding one of the Lowest Landau Level(LLL)
structure. We discuss these two cases separately in what follows.

\subsection{Trapped rotating fermions}

When the particle number in the two hyperfine states are equal the system
is unpolarized. Since the interaction strength is symmetric, the turning
points are therefore the same. This situation changes only if there is a
spontaneously broken symmetry in the ground state leading towards phase
separation. 

For the unpolarised case the spin density $\ti{s}(\vec x)$
vanishes everywhere. Adding eqs.(\ref{rho1}) and (\ref{rho2}) 
we obtain  
\beq 
2 \pi \nabla^2 \ti{\rho} + 4 \ti{\Omega}^2 -2 \pi \ti{g}\ti{\rho} 
+\frac{1}{\ti{b}^2}(\ti{\mu}^+ - 2 \pi\ti{\rho} -x^2 \ti{\Omega}^2) = 0 
\label{rho} 
\eeq
The same equation may also be obtained by minimising the energy functional
given in the previous section, see eq.(\ref{E}) where we simply considered
a system with spin degeneracy of 2 as in the unpolarised case. 

The solution of the above equation may be written in terms of the 
modified Bessel function, namely
\beq 
2\pi \ti{\rho} = AI_0(\eta x) + \frac{1}{\eta^2
\tilde{b}^{2}}[\tilde{\mu}^{+} + \frac{4
\ti{\Omega}^{2}\ti{b}^{2}\ti{g}}{\eta^{2}} - x^{2} \ti{\Omega}^{2}]  ,
\label {srho} 
\eeq 
where 
\beq 
\eta^2 = (\ti{g} + \frac{1}{\ti{b}^2}) 
\eeq
The solution for the density has three unknown parameters, the turning
point $x_0$, the overall normalisation $A$ and the chemical potential
$\tilde{\mu}^{+}$. In the following we outline how these may be determined 
from
the known constraints. The first constraint arises from the fact that at
the classical turning point the density vanishes. We may use this 
condition to determine the chemical potential:
\beq
\ti{\mu}^{+} = -\ti{b}^{2} \eta^2
AI_0(\eta x_0) + \ti{\Omega}^{2} x_{0}^{2} - \frac{4\ti{g}\ti{\Omega}^{2}
 \ti{b}^{2}}{\eta^2} 
\label{chem} 
\eeq
This gives $\ti{\mu}^{+}$ in terms of the other two unknowns $A$ and
$x_0$. Substituing this we get 
\beq
2 \pi\ti{\rho} = AI_0(\eta x_0)[\frac{I_0(\eta x)}{I_0(\eta x_0)} - 1] +
\frac{\ti{\Omega}^{2}}{\ti{b}^{2}\eta^{2}}[x_{0}^{2} - x^2] 
\eeq
Interestingly, the first term in the above expression corresponds to the
density obtained in the case of logarithmic interaction and the second
term corresponds to the density obtained in the case of contact
interaction if the coefficients are properly adjusted. These different
limits were derived earlier in Ref.\cite{ghosh} separately for these two
interactions. 
 
The second constraint is given by the continuity of the Hatree potential
and its derivative which is the electric field experienced by the charge
distribution at the turning point. Adding eq. (\ref{pot1}) and eq.
(\ref{pot2}) we have
\beq 
\nabla^2 \ti{V}_{H+} = -2 \pi \ti{g}\ti{\rho}
+\frac{\ti{V}_{H+}}{\ti{b}^2},
\label{HATREE} 
\eeq
where $\ti{V}_{H+}$ corresponds to the Hatree potential
with 
$$\ti{V}_{H+} = \ti{V}_{H1}+\ti{V}_{H2} $$
yIn the region $x\ge x_0$ the equation (\ref{HATREE}) takes the form
\beq 
\nabla^2 \ti{V}_{H+}= \frac{\ti{V}_{H+}}{\ti{b}^2} 
\eeq 
since the density is zero beyond the turning point. The solution is  
given by 
\beq 
\ti{V}_{H+}=\frac{A_1K_0(x/\ti{b})}{\ti{b}^2} ,
\eeq 
where $A_1$ is an overall normalisation which is not needed in what 
follows. 
The electric field is just the negative of the gradient of this potential
$\sc{E} = -\nabla V$.
We define the ratio of the potential to the field as follows
\beq \frac{\ti{V}_{H+}(x_0)}{{\sc E}(x_0)} =
\ti{b}K_0(x_0/b')/K_1(x_0/b') 
\label{POTFIELD1} 
\eeq
From  eq. (\ref{tf1}) and (\ref{tf2}) we have for $x \le x_0$ 
\beq 
\ti{V}_{H+} =\ti{\mu}_{+} - \ti{\Omega}^2 x^2 - \ti{\rho}
\eeq
Again taking the ratio of the potential and the electric field we have
\beq 
\frac{\ti{V}_{H+}(x_0)}{{\sc E}(x_0)} = -\frac{ \mu -\Omega^2x_0^2 
- \ti{\rho}(x_0)}{2\Omega^2 x_0 +(\ti{\rho}(x))'|_{x=x_0}} 
\label{POTFIELD2} 
\eeq
Now demanding the continuity of the potential and the field across the 
turning point we have from eqs.(\ref{POTFIELD1}) and (\ref{POTFIELD2})
\beq 
AI_0(\eta x_0)=-\frac{2 \ti{\Omega}^2 \ti{g}\ti{b}^{2}}{\eta^2}
\frac{(2 + \frac{x_0 K_0(x_0/\ti{b})}{\ti{b} K_1(x_0/\ti{b})})}
{(\ti{b}^2 \eta^2 +\ti{b}\eta [\frac{(I_1(\eta
x_0)K_0(\frac{x_0}{\ti{b}})}{I_0(\eta x_0)K_1(\frac{x_0}{\ti{b}})}])}
\eeq 
which now gives the relation between $A$ and the turning point.
Finally the turning point, and hence all the undetermined parameters are 
determined using the fact that the number of particles in the system is 
fixed. That is,
\beq 
N=AI_0(\eta x_0)[\frac{x_0I_1(\eta x_0)}{\eta I_0(\eta x_0)}
-\frac{x_0^2}{2}]
+\frac{1}{4}\frac{\ti{\Omega}^2 x_0^4}{\ti{b}^2 \eta^2} 
\label{NOP1}\eeq
Note that the turning point increases with N in a nonlinear fashion given
by the above relation. For  large number of particles,however, the
turning point is given by the following asymptotic relation
\beq 
N  \approx \frac{1}{4}\frac{\tilde{\Omega}^2 x_0^4}{\tilde{b}^2
\eta^2}
\label{asymp1} \eeq
All the other relevant quantities defining the system may now be written 
down. The exact expression for the  total angular momentum is given by
\beq 
\frac{L}{\hbar}
= AI_0(\eta x_0)\frac{\omega}{\omega_0}[(\frac{x_0}{\eta}(x_0^2
+\frac{4}{\eta^2})\frac{I_1(\eta x_0)}{I_0(\eta x_0)} 
-\frac{2 x_0^2}{\eta^2} - \frac{x_0^4}{4}] 
+\frac{\omega}{\omega_0}\frac{\ti{\Omega}^2}{\ti{b}^2 \eta^2}
\frac{x_0^6}{12} 
\eeq

The Energy is again given by the expression
\beq 
\frac{E}{\hbar \omega_0} = \frac{\ti{\mu}^{+}N}{2}
+\frac{1}{4}\frac{\omega_0^2 - \omega^2}{\omega \omega_0}\frac{L}{\hbar} 
\eeq

It is easy see that all the above expressions above can be written as a
sum of contributions coming from the purely contact interaction and the
purely logarithmic term (\cite{ghosh}), two extreme limits of the screened
Coulomb interactions. However this is not valid for the interaction energy
since it is bilinear in the density.  This may be seen from the analytic
expression for the interaction energy 
\bea \frac{E_{int}}{h \omega_0} &=&
\frac{E}{h \omega_0} - \pi^2\int \ti{\rho}^2 xdx -
\frac{1}{2}\frac{\ti{\Omega}^{2}\omega_0}{\omega} \frac{L}{h} \nonumber \\ 
&=&
\frac{E}{\hbar \omega} -\frac{1}{4}(I_1+I_2+I_3) -
\frac{1}{2}\frac{\ti{\Omega}^{2}\omega_0}{\omega} \frac{L}{h}, \eea 
where 
\bea
I_1&=&[(AI_0(\eta x_0)^2[\frac{x_0^2}{2}[1-\frac{I_1^2(\eta
x_0)}{I_0^2(\eta
x_0)}] -2x_0\frac{I_1(\eta x_0)}{I_0(\eta x_0)}+\frac{x_0^2}{2}]] \\
I_2 &= & AI_0(\eta x_0)\frac{\ti{\Omega}^{2}}{\tilde{b}^2
\eta^2}[2\frac{x_0^2}{\eta^2}[1-\frac{2}{x_0 \eta}
\frac{I_1(\eta x_0)}{I_0(\eta x_0)} -
\frac{x_0^2\eta^2}{8}]] \\
I_3&=&\frac{1}{4}(\frac{\ti{\Omega}}{\tilde{b}\eta})^4\frac{x_0^6}{6} 
\eea 

While the system is analytically solvable as shown above, the actual
behaviour is not often easy to see.  We illustrate the results for
reasonably large systems with number of particles $N=50,100, 1000$.  The
numerical calculation is done in two different scenarios. In the first
case we have kept the particle number $N$, the range $\tilde{b}$ and
coupling strength $\tilde{g}$ fixed and vary the external frequency of
rotation.  In the second case we have calculated the relevant quantities
as a function of the quantity $\ti{g}\ti{b}^2$ where all the other quantities
except the range $\ti{b}$ are kept
fixed. 

In Fig.2 we show the turning point $x_0$ as a function of the frequency
$\omega$ measured in units of $\omega_0$. Note that Eq.(\ref{NOP1}) is
highly nonlinear. In order to estimate the turning point we use the value
given by the asymptotic result as an upperbound and use this value to
iterate the exact equation to obtain the exact value of $x_0$. It is easy
to see that the system size increases with rotational frequency. All other
quantities are determined once the turning point is known. The system 
size increases with rotational frequency rather slowly as expected. The 
size variation with respect to the number of particles is approximately 
given by the asymptotic formula.

The angular momentum per particle also follows a similar behaviour.
as shown in Fig. 3. The ratio of the energy of the system measured in 
terms of the ground state energy is shown in Fig.4. Shown also in this 
figure is the interaction energy for $N=1000$. At low frequencies the 
interaction contributes substantially to the total energy. However, at 
large frequencies the interaction contributes little, and most of the 
total energy is simply the energy of rotation which is proportional to 
the angular momentum $L$. Following the analogy with rotating nuclei, 
this is usually referred to as the Yrast region \cite{Bohr} and has been a 
subject of intense study in recent times\cite{Mot1}. 

As already pointed out, the screened Coulomb interaction can mimick
different types of two body interaction depending on the range. 
We study
the system for $N=1000$ at a fixed frequency of $\omega=0.25 \omega_0$, in
Figs. 5,6,7 where we have plotted the turning point, the angular momentum
per particle and the total energy respectively as 
a function of  $\ti{g}\ti{b}^2$. 
The quantity $\ti{g}{b}^2$ here is analogous  
to the coupling strength in the pseudopotential of a  
three dimensional Bose condensate,  $g_{3D}=\frac{4 \pi \hbar^2 a}{m}$, 
where
$a$ is the s-wave scattering
length. Note that the system of fermions, for these purposes, may be
regarded as a hardcore bosons.
This three dimensional coupling constant is related to the 
two dimensional coupling constant by the formula 
(\cite{Castin}) $g_{2D}=g_{3D}\sqrt{\frac{m\omega_z}{h}}$.
We have estimated this constant using the parameters for $^{40}K$
and chosen the appropriate range of $\ti{g}\ti{b}^2$
(see ref. \cite{Tarun} for such an estimate). 
 From these figures, it is
easy to see that within the Thomas Fermi approximation the effect of
change of range is considerable even at moderate rotational frequency as
above.  In particular in Fig.7, we see that both the total and
interaction energy increase with the range.  

It should be mentioned that the analysis can also be extended to the
partially polarised system where the number  of particles in different
components are not same . Formally one can write the
solutions in closed
form which however are more difficult to evaluate since the turning point
for each component density will be different. This case will be considered
in detail in another paper\cite{future}. 

\subsection{Rapidly Rotating fermions}

The above TF analysis is valid when the trapping frequency is larger than
the rotational frequency. The effect of rotations is to reduce the
effective trapping frequency and in the absence of interactions this leads
to bunching of single particle levels.  There are thus many similarities
between the system of rotating fermions and the fermions in the presence
of an external magnetic field along z-direction. The rotation frequency to
plays a role analogous to the cyclotron frequency. We exploit this
similarity in what follows. 

Noninteracting Hamiltonian for a rapidly rotating system is given by,
\beq
H_{0} = \sum_{i} \frac{\hat{p}_{i}^{2}}{2 m} + \frac{1}{2}m \omega_0^{2} 
r_{i}^{2}
-\omega \hat{L}_{i},
\eeq
where as before $\omega$ is the external rotation frequency. The single 
particle energy levels are given by,
\beq
E_{n,l} = (2n + |l| + 1)\hbar \omega_{0} - \hbar \omega l,
\eeq
where $n$ is the radial quantum number. This spectra is the same as that
of a cranked harmonic oscillator with two frequencies $\omega_0 + \omega$
and $\omega_0 - \omega$ (for details see Ref.  \cite{bhaduri}). The Landau
level like structure starts to form when $(\omega_0 - \omega) << (\omega_0
+ \omega)$ and the mixing between the different levels is supressed. In
particular when $ \omega_0 - \omega = 0$, the spectrum becomes identical
to that of a set of highly degenerate discrete Landau levels. In the other
limit
when the external rotation frequency is not too high and $\mu >> (\omega_0 
+ \omega)$, so that many levels are occupied, the system is described by 
Thomas-Fermi approximation.  

In the lowest Landau level(LLL) the particles have the radial quantum
number $n = 0$ and the angular momentum $l \ge0$.  The single particle
wave functions and energies in LLL are given by,
\bea
\psi_{l} & = & \frac{1}{\sqrt{\pi l_{0}^2 l!}} (\frac{z}{l_{0}})^{l} e^{-r^{2}/2
l_{0}^{2}}, \label{wf} \\
E_{l} & = & (\omega_{0} - \omega) \hbar l + \hbar \omega_{0},
\label{ee} \eea
where $l_{0}^2 = \frac{\hbar}{m \omega_{0}}$, and $z = x + i y$.

The density of the non-interacting spin polarised electrons is given by,
\bea
\rho_{0} & = & \frac{1}{\pi l_{0}^{2}} \sum_{l=0}^{N} \frac{1}{l!}
(z/l_{0})^l e^{-r^2
/l_{0}^2}, \\
& \approx & \frac{1}{\pi l_{0}^{2}} \theta (R - r),
\eea
where, $R$ is the radius of the system for large number of particles. 
In this limit the particles form a flat droplet like state.

We now consider the interacting two component unpolarised fermions in 
the ``lowest Landau band'' when the rotation frequency is almost equal 
to the external trap frequency. The interaction is taken to be the screened 
Coulomb interaction defined earlier. Since TF method can not be applied 
directly, we use the standard
variational method for determining the ground state properties.
We construct a local density
functional by taking the expectation value of the quantum mechanical 
many particle Hamiltonian with the variational state given above,
neglecting the mixing between the ground and excited states.
The beauty of this method is that all quantities of 
physical interest may be derived analytically. Though this method
also does not goes beyond the mean field approximation , it
performs better compared to the bare TF method since the variational 
ansatz does account for quantum mechanical effects. 
  
The many body variational wave function for each component may be written 
as, 
\beq
\psi(\{z_{i}\}) = f(\{z_{i}\}) e^{-\sum_{i} z_{i}\overline{z}_{i}/2 
l_{0}^{2}}. 
\eeq
Kinetic energy of the system  is then given by,
\beq
< T >  =  <\psi |-\frac{\hbar^{2}}{2 m} (4\frac{\partial}{\partial
 \overline{z}}\frac{\partial}{\partial z})| \psi>
=  \frac{1}{2} m \omega_{0}^{2} \int r^{2} \rho(r) d^{2} r.
\eeq
Note that the kinetic energy density is linear in $\rho$, that is, 
$T[\rho] = \frac{1}{2}m \omega_{0}^{2} r^{2} \rho(r)$,where as in the 
standard TF case it is quadratic.
The total angular momentum of the two-dimensional system is given by,
\beq
< L >  =  -\hbar \int [\psi^{*} \overline{z} \frac{\partial \psi}{\partial
\overline{z}} - \frac{\partial \psi^{*}}{\partial z} z \psi] - \hbar
\int \psi^{*} 
\psi  
=  \int d^{2}r [m \omega_{0} r^{2} - \hbar] \rho(r).
\eeq

Using these, the energy density functional of two component fermi gas can be 
written as, 
\beq
E[\rho_{1},\rho_{2}]  =  T[\rho_{1}] + T[\rho_{2}] + V[\rho_{1}] + V[\rho_{2}] -
\Omega (L[\rho_{1}] + L[\rho_{2}]) + E_{int}[\rho_{1}, \rho_{2}].
\eeq
This energy functional is similar to the magnetic field Thomas Fermi
method developed by Lieb et al \cite{lieb}.
For the unpolarised system, $\rho_{1} = \rho_{2}$.  The free energy of 
the system may be written interms of $\rho = \rho_1+\rho_2$ alone. Using 
the expression for kinetic energy density and scaling to dimensionless 
variables as before, the free energy of the two component system is given by
\beq
\frac{F[\rho]}{\hbar \omega_{0}} = (1 - \ti{\omega})\int
d^{2}x x^{2} \ti{\rho}
+ \frac{1}{4}
\int d^{2}x_{1} d^{2}x_{2} \ti{\rho}(x_{1}) \ti{V}_{12}(|\vec{x_{1}} -
\vec{x_{2}}|)
\ti{\rho}(x_{2}) - (\ti{\mu} -\ti{\omega})\int d^{2}x \ti{\rho},
\eeq
where $\ti{\omega} = \frac{\omega}{\omega_0}$.

Minimising the free energy for a fixed number of particles gives the 
following integral equation for the density:
\beq
(1 - \ti{\omega}) x^{2} + \frac{1}{2} 
\int d^{2}x'\ti{V}_{12}(|\vec{x} -
\vec{x'}|) \rho(x')=(\ti{\mu}-\ti{\omega}),
\eeq
As noted before, the screened Coulomb potential  satisfies,
the following differential equation,
\beq
\nabla^{2} \ti{V}_{H} = -2 \pi \ti{g} \ti{\rho} +
\frac{\ti{V}_{H}}{\ti{b}^{2}}.
\eeq

While the equation system looks complicated, the solution for the 
circularly symmetric case is rather simple and is given by,
\beq
\rho = \frac{\alpha}{\pi}(\tilde{\mu}_c - x^{2}),
\eeq
where
\bea
\alpha &=& \frac{1 - \ti{\omega}}{\ti{g} \ti{b}^{2}},\\
\tilde{\mu}_c &=& \frac{\ti{\mu} - \ti{\omega}}{1 - \ti{\omega}} + 4
\ti{b}^{2}.
\eea
One can easily recognise that the solution bears a resemblence to that of 
the lowest Landau level droplet for the rapidly rotating fermions.

The size of the droplet ($R$) may be obtained by normalising the above
density 
to yield the total number of particles $N$  which is
fixed. We 
have 
\bea
\tilde{\mu}_c &=& R^{2}\nonumber\\
R &=& (2N/\alpha)^{1/4}.
\eea

Now we are in a position to compute all quantities of interest relevant 
to the rapidly rotating two-component system.
Angular momentum of the 
system is given by
\beq
L =  \frac{<L>}{\hbar} = \frac{1}{6} \alpha (2N/\alpha)^{3/2} - N.
\label{rram} 
\eeq
For a given angular momentum this equation determines the parameter 
$\alpha$ interms of the fixed parameters $L$ and $N$. That is,
\beq
\alpha = \frac{2}{9} \frac{N^{3}}{(L + N)^{2}}.
\label{alpha} \eeq
The interaction energy is then given by,
\beq
E_{int}(L,N,\ti{g},\ti{b}^{2}) =  \frac{2}{9}\frac{N^{3} \ti{g}
\ti{b}^{2}}{(L +N)^{2}}
[L + N(1 - 2\ti{b}^{2})].
\label{rrint} 
\eeq
The total energy in the rotating frame is given by,
\bea E(L,N,\ti{g},\ti{b}^{2})&=&
(1-\ti{\omega})\int d^{2}x x^2 \ti{\rho}  +\ti{\omega}\int d^{2}x
\ti{\rho} + E_{int} \nonumber \\ 
&=&  N + \frac{2}{9}\frac{N^{3} \ti{g}
\ti{b}^{2}}{(L +N)^{2}}[2L + N(1 - 2\ti{b}^{2})] \label{rrtot}
\eea
 Variation of the total energy with external rotation frequency is
given by,
\beq
E(\tilde{\omega},N,\tilde{g},\tilde{b}^{2}) = N + \sqrt{\tilde{g}
\tilde{b}^{2} (1 - \tilde{\omega})} (2N)^{3/2} - (1 -
\tilde{\omega})N(1+ 2 \tilde{b}^{2}).
\eeq 
Since $\alpha \propto (1 -\ti{\omega})$, the angular momentum increases as
the rotational frequency approaches the confinement frequency. In fact
when $\omega=\omega_0$, the total angular momentum blows up, and the
system resembles the LLL droplet. The total energy is given entirely by
the energy of non-interacting LLL system since the interaction goes to zero. 
Total energy goes to the zero point energy as $\sim \sqrt{1 -
\tilde{\omega}}$.
\section{Summary}

We have discussed a Thomas-Fermi model for a rotating two-component
interacting Fermi vapour in two dimensions. We believe that the
interaction, the screened Coulomb, is realistic with interesting limits as
the range is varied. The results correponding to short range and
logarithmic interactions may be obtained with suitable approximations.
More importantly the system is analytically solvable within the TF mean
field approach. We have neglected the diagonal component of the
interaction including the dominant s-wave off diagonal interaction.  We
have first considered the case when the trap frequency is larger than the
rotational frequency since one can use the TF method ab initio.  However
when the rotational frequency is much larger than the trap frequency we
first projected the system on to the lowest degenerate level and used a TF
like approach to derive the analytical result. It should be noted that, in
general, it is rather hard and near impossible to obtain accurate
results for large number of particles using exact diagonalisation. In 
contrast, the TF method is solvable in a realistic case as shown and 
provides a reasonably good approximation. An important drawback in the 
present analysis is that we have not included the exchange corrections. 
This however is obviated if the number of particles in the system is 
large and sufficiently dilute.

To put the utility of the method in perspective, unlike in the case of
superfluid bosons, the system of fermions has a rigid body rotation. A
direct measurement of the angular momentum in a system of bosons is very
difficult. But in the case of fermions total angular momentum can be
estimated just from a measurement of the moment of inertia. It is valid
for
low rotation frequencies, where the usual TF method is good. The radius of
the cloud can be measured easily by observing the diffusion of the gas (by
switching off the trap potential). In the rapidly rotating case, however,
the total angular momentum is not proportional to $<r^2> $ due to the
quantum effects being large and a direct measurement of the angular
momentum would be difficult. The system is however more interesting
because of its similarity to the LLL problem. It would be interesting to
see if the system exhibits quasiparticle (quasi hole )like excitations
which is similar to the vortex excitations in the corresponding Bose
systems.

One of us (S.S) acknowledge financial support from Minist\`ere de la
Recherche et de la Technologie.
LKB is a unit\'e de recherche de l'Ecole normale sup\'erieure et de
l'Universit\'e Pierre et Marie Curie, associ\'ee au CNRS.

\newpage

\begin{figure}
\label{fig1}
\caption{The spatial variation of three type of potentials 
namely $\frac{1}{x}$ (dashed),~$-log|x|$ (dotted) and the screened Coulomb 
$K_0(x)$ (solid line)(where we have set $\tilde{b}=1$) is shown.}
\end{figure}

\begin{figure}
\label{fig2}
\caption{The plot of turning point of the rotating fermions $x_0$ as a 
function of the external frequency $\omega$. The unit of $\omega$ is 
the confinement frequency $\omega_0$ and the unit of length is  
$l_0 = \sqrt{\frac{\hbar}{m\omega_0}}$. The three curves correspond to
$N=50$ (solid line),$N=100$ (short-dashed) and $N=1000$ (long -dashed).
 For all the curves plotted we have chosen $\tilde{b}=0.15$ and
$\tilde{g}=0.1$.}
\end{figure}

\begin{figure}
\label{fig3}
\caption{The plot of angular momentum per particle  of the rotating
fermions $L/N$ as a
function of the external frequency $\omega$. The unit of $\omega$ is
the confinement frequency $\omega_0$ and the unit of $L$ is
$\hbar$.The three curves correspond to
$N=50$ (solid line),$N=100$ (short-dashed) and $N=1000$ (long -dashed).
Again  for all the curves plotted $\tilde{b}=0.15$ and $\tilde{g}=0.1$.}
\end{figure}

\begin{figure}
\label{fig4}
\caption{The plot of energy  of the rotating fermions as a function of
external frequency $\omega$.The unit of $\omega$ is same as that in 
Fig.3. The  energies are normalised to the  
$\omega =0$  case. The upper curve 
with dashed line shows the 
variation of the total
energy and  the lower curve (solid line)  
shows the variation of the interaction energy.
The number of particles is   $N=1000$ and  $\tilde{b}=0.15$, $
\tilde{g}=0.1$}
\end{figure} 

\begin{figure}
\label{fig5}
\caption{The plot of the turning point of the rotating fermions as a
function of $\tilde{g}\tilde{b}^2$ when $\tilde{b}$ is varied.
The number of particles $N$, $\tilde{g}$ 
and the frequency $\omega$  (in units of 
$\omega_0$) are respectively given by 
$1000, 0.1, 0.25$.}
\end{figure}
 
\begin{figure}
\label{fig6}
\caption{The plot of angular momentum per particle $L/N$
as a function of  $\tilde{g}\tilde{b}^2$ when 
$\tilde{b}$ is varied. The unit of 
angular momentum and length is again $\hbar$ and $l_0$ respectively.
The number of particles $N$, $\tilde{g}$
and the frequency $\omega$  (in units of
$\omega_0$)  are respectively set at
$1000, 0.1, 0.25$.}
\end{figure}

\begin{figure}
\label{fig7}
\caption{The plot of energy as a function of  $\tilde{g}\tilde{b}^2$
when $\ti{b}$ is varied.
The continuous curve corresponds to the total energy where the 
dashed curve corresponds to the interaction energy alone. 
The number of particles $N$, $\tilde{g}$                      
and the frequency $\omega$  (in units of
$\omega_0$) are given by
$1000, 0.1, 0.25$}
\end{figure}

\end{document}